\documentclass[9.5pt,journal,compsoc,peerreview]{IEEEtran}

\ifCLASSOPTIONcompsoc
  \usepackage[nocompress]{cite}
\else
  \usepackage{cite}
\fi
\usepackage{microtype}                 
\PassOptionsToPackage{warn}{textcomp}  
\usepackage{textcomp}                  
\usepackage{mathptmx}                  
\usepackage{times}                     
\usepackage{cite}                      
\usepackage{tabu}                      
\usepackage{booktabs}                  
\usepackage{soul}
\setstcolor{red}

\usepackage{graphicx}        
\usepackage{url}
\usepackage{hyperref}
\usepackage{amsmath,amssymb}
\usepackage{subfigure}
\usepackage{float}
\usepackage{multirow}
\usepackage[table,xcdraw]{xcolor}

\usepackage{fancyhdr,graphicx,amsmath,amssymb}
\usepackage[ruled,vlined]{algorithm2e}

\newcommand{\blue}[1]{\textcolor{black}{#1}}
\newcommand{\red}[1]{\textcolor{pink}{\st{}}}

\begin{document}
%
\title{Modeling Data-Driven Dominance Traits for Virtual Characters using Gait Analysis}
%
%
%
%

\author{Tanmay~Randhavane, Aniket~Bera, Emily Kubin, Kurt Gray, Dinesh Manocha
\IEEEcompsocitemizethanks{\IEEEcompsocthanksitem T. Randhavane and A. Bera are with the Department of Computer Science, University of North Carolina, Chapel Hill, NC, 27514. \protect\\
E-mail: tanmay@cs.unc.edu
\IEEEcompsocthanksitem E. Kubin is with the Department of  Social Psychology, Tilburg University, Tilburg, Netherlands.
\IEEEcompsocthanksitem K. Gray is with the Department of Psychology and Neuroscience, University of North Carolina, Chapel Hill, NC, 27514.
\IEEEcompsocthanksitem D. Manocha is with the Department of Computer Science and Electrical \& Computer Engineering, University of Maryland, College Park, MD, 20740.}
}

%
%

\markboth{IEEE Transactions on Visualization and Computer Graphics}%
{Randhavane \MakeLowercase{\textit{et al.}}: Modeling Data-Driven Dominance Traits for Virtual Characters using Gait Analysis}
%



\IEEEtitleabstractindextext{%
\begin{abstract}
We present a data-driven algorithm for generating gaits of virtual characters with varying dominance traits. Our formulation utilizes a user study to establish a data-driven dominance mapping between gaits and dominance labels. We use our dominance mapping to generate walking gaits for virtual characters that exhibit a variety of dominance traits while interacting with the user. Furthermore, we extract gait features based on known criteria in visual perception and psychology literature that can be used to identify the dominance levels of any walking gait. We validate our mapping and the perceived dominance traits by a second user study in an immersive virtual environment.  Our gait dominance classification algorithm can classify the dominance traits of gaits with \blue{\texttildelow 73\%} accuracy. We also present an application of our approach that simulates interpersonal relationships between virtual characters. To the best of our knowledge, ours is the first practical approach to classifying gait dominance and generate dominance traits in virtual characters.
\end{abstract}

}

\maketitle

\IEEEdisplaynontitleabstractindextext

%
\IEEEpeerreviewmaketitle

\IEEEraisesectionheading{\section{Introduction}\label{sec:introduction}}
There is considerable interest in simulating human-like virtual characters. The set of applications for these simulations includes training, social VR, gaming, virtual crowds, VR therapy, urban modeling, etc. There are many challenges involved in generating the appearance, movements, and plausible behaviors of such virtual human-like characters.  The behaviors to be generated include verbal and non-verbal behaviors. Furthermore, many applications need to simulate virtual characters with varied personalities and traits. 


Prior studies and evaluations in psychology and virtual environments have shown that some of the components of pedestrian movement, including joint positions and orientations, are important for realistic human perception~\cite{Ennis2011SceneContext,Pelechano2008Presence}. As a result, there is considerable recent work on generating plausible trajectories or movements of virtual characters. In the physical world, humans are known to be adept at using many visual cues, including subtle cues, to make impressions of or social judgments about others. As a result, there has been some work on simulating and evaluating the benefits of non-verbal behaviors like gaze, gestures, and gaits~\cite{bailenson2005transformed,pedvr}.
 
In this paper, we focus on analyzing gait features in real-world scenes and using them to generate gaits of virtual characters that can display a spectrum of dominance traits. A person's gait or style of walking is a unique feature of their overall movement. Gaits have been shown to be an effective biometric cue for visual identification~\cite{wang2003silhouette} or recognition of biological motion~\cite{beardsworth1981ability}. Previous studies have shown that humans can convey and perceive much information including sex differences~\cite{bruening2015sex}, emotions~\cite{kleinsmith2013affective}, moods,  and personality traits from gaits. Moreover, walking style also affects what people tend to remember~\cite{michalak2015we}. Overall, non-verbal cues like gaits and the style of walking can guide the perception of emotions~\cite{roether2009critical,atkinson2004emotion}, moods, and personality traits of characters~\cite{heberlein2004cortical}, including dominance and openness.

\begin{figure}[t]
    \includegraphics[width=1\linewidth]{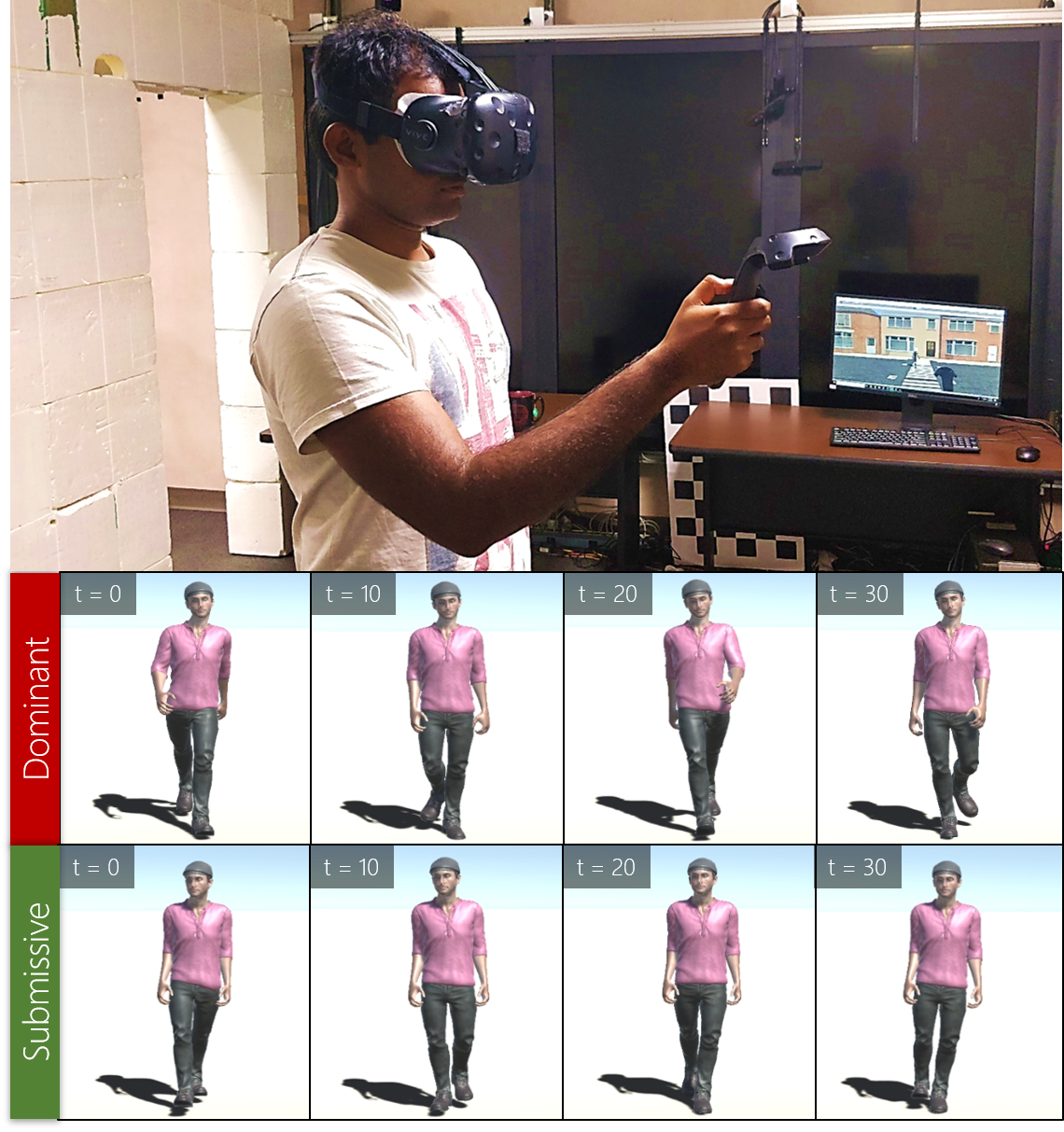}
    \vspace{-18pt}
    \caption{\textbf{Modeling Dominance Traits for Virtual Characters:} Our approach learns dominance traits from motion-captured gaits and computes a data-driven dominance mapping. We use this mapping to \blue{interactively} generate virtual characters with different dominance traits (below) for an immersed user (top). According to the psychology literature, more leg and hand movement and erect posture indicates a dominant gait, whereas slumped posture and less leg and hand movement indicates a submissive gait.}
    \label{fig:frontCover}
\end{figure}

In terms of modeling any human-like behavior of a virtual character, it is important to define the behavior and understand the factors that influence that behavior. Previous research has formulated \textit{dominance behavior} as the behavior directed towards the control of another through implied or actual threat ~\cite{ridgeway1987nonverbal}.  Dominance can be defined as a personality trait involving the motive to control others or as some perception of oneself as controlling others. The dominance behavior is manifested through various verbal and non-verbal cues. We focus on non-verbal cues associated with walking gaits to generate virtual characters with dominance traits. 
This includes the study of behaviors like postural erectness, postural openness, slumped shoulders, etc., which affect the perception of dominance. We also use variations in walking speed where characters with fast-moving gaits are perceived as more dominant than characters with slow-moving gaits.
Our work is also inspired by prior work on the visual perception of human gaits. In particular, Johansson~\cite{johansson1973visual} showed that a few bright spots, which are used to describe the motions of the main joints of humans, are sufficient to evoke a compelling impression of human activity. Humans can perceive a considerable amount of information from watching other humans' gaits. Therefore, dominance traits of virtual characters can also be modeled using their walking gaits.

\begin{figure*}[t]
  \centering
  \includegraphics[width =\linewidth]{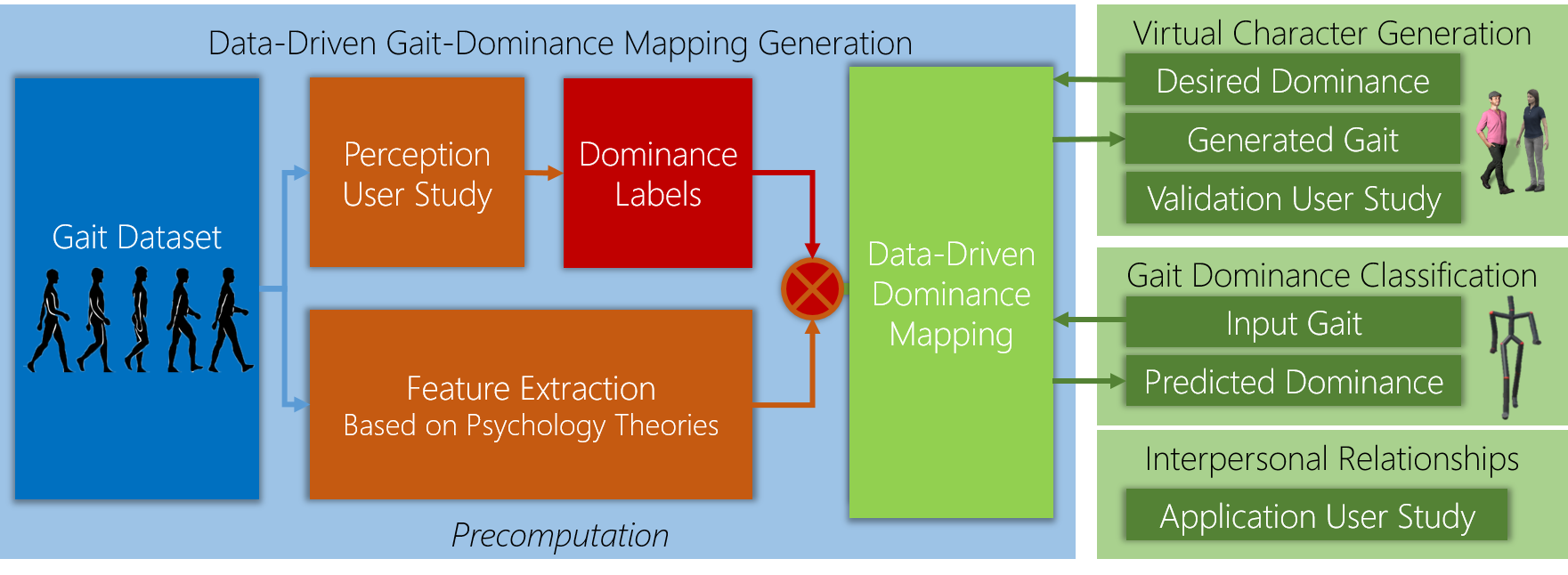}
  \vspace{-18pt}
  \caption{\textbf{Overview:} We highlight the various components of our approach. We start with a gait dataset and perform a perception user study to generate different dominance labels. We also compute different gait-based features representing dominance according to prior work in psychology. Finally, we compute a data-driven mapping between the dominance labels and the features. To generate a character with a specific dominance trait, we use the corresponding gait at real-time using our dominance mapping. Given new gaits, we compute their features and predict the dominance labels. Our approach can be used to simulate interpersonal relationships relating to dominance between virtual characters.}
  \label{fig:overview}
\end{figure*}

\textbf{Main Results:} We present a data-driven model of dominance traits of virtual characters based on gait analysis \blue{(Figure~\ref{fig:frontCover})}. Our formulation is based on using the results of a user study on gait datasets~\cite{CMUGait,narang2017motion,ma2006motion} to generate dominance labels for the walking gaits. We use these annotated walking gaits to generate virtual characters with different dominance traits. Based on prior work in visual perception on non-verbal cues related to perceived dominance~\cite{furley2012nonverbal, weisfeld1982erectness,curran2016methods,richards1991perceptions,eisenburg1937expressive}, we extract appropriate features from human gaits. Next, we classify these features into dominance labels using Support Vector Machines trained on the annotated dataset. Overall, our method can be used to identify the dominance of any gait. We also present an application of our approach to simulate interpersonal social relationships relating to dominance between virtual characters.


An overview of our approach is given in Figure 2, where we highlight offline dominance mapping generation and its application to runtime virtual character generation and gait dominance classification algorithms. 
Our cross-validation results show that our method can be used to classify the dominance traits of any new gait with \texttildelow73\% accuracy.

Our contributions include three user studies to model and validate dominance traits:
\begin{itemize}
    \item Perception User Study: In this web-based study, we obtain dominance labels for gaits that are used to establish a gait dominance mapping (Section~\ref{sec:perceptionStudy}).
    \item Validation User Study: In this immersive study, we validate our approach to generating virtual characters with different dominance traits in a virtual environment. We observe statistically significant differences between the dominance traits of virtual characters created using our approach.
    Results of this extensive study indicate that our data-driven approach can be used to generate virtual characters with the desired levels of dominant and submissive behaviors (Section~\ref{sec:validationStudy}).
    \item Application User Study: In this immersive study, we present an application for simulating interpersonal social relationships between virtual characters (Section~\ref{sec:applicationStudy}). The results of this study indicate that our approach can be used to simulate the interpersonal social relationships relating to dominance between virtual characters.
\end{itemize}


Overall, our approach to model dominance has the following benefits:  

\noindent {\bf 1. Interactive Performance:} Our approach can be used to interactively generate virtual characters that exhibit a range of dominance traits.

\noindent {\bf 2. Compatibility:} Our approach is orthogonal to different methods used for local navigation or the generation of other personality traits, gaze, emotions, moods, etc.

\noindent {\bf 3. Generalizability:} Our approach can be used to identify the dominance levels of any new gait.

\noindent {\bf 4. Interpersonal social Relationships:} Our approach can be used to simulate interpersonal social relationships relating to dominance among virtual characters.


The rest of the paper is organized as follows. In Section 2, we review the related work in the fields of behavior modeling and gait analysis. In Section 3, we describe our \blue{perception user study}. In Section 4, we describe our gait dominance classification algorithm. We present the details of our \blue{validation user study} in Section 5 and our classification results in Section 6. In Section 7, we present an application of our approach that simulates interpersonal social relationships between virtual characters and the \blue{application user study. We conclude with the limitations and future opportunities of our work in Section 8.}

\section{Related Work}
\label{sec:RelatedWork}

In this section, we give a brief overview of prior work on behavior modeling of virtual characters and gait analysis.

\subsection{Gait Analysis}
There is considerable work on automatic recognition of emotions and personalities from body expressions such as gaits. Most works use a feature-based approach where the features are either extracted using purely statistical techniques or are inspired from psychological studies. Some approaches focus on specific activities such as dancing~\cite{raptis2011real}, knocking~\cite{gross2010methodology}, walking~\cite{karg2010recognition}, games~\cite{kleinsmith2011automatic}, etc., whereas other approaches use a more generalized approach~\cite{crenn2016body,wang2016adaptive}. Some techniques combine both facial and body expressions~\cite{pollick2004combining,van2007body,meeren2005rapid,tuminello2011face,willis2011judging,clavel2009combining,gunes2006observer,kapoor2007automatic}. Janssen et al.~\cite{janssen2008recognition} use neural networks to identify emotions from gaits. They observe that the differences between the gaits of individuals were much larger than the differences between emotions. Other approaches find emotions expressed in gaits with the help of neutral expressions~\cite{roether2009critical,crenn2017toward}.  Studies have shown that both posture and movement cues are important for the perception of emotion and personality~\cite{atkinson2007evidence,roether2009critical}. Gaits have also been used for activity recognition~\cite{presti20163d,aggarwal2014human}. Although there is a lot of work on modeling emotion and other personality characteristics~\cite{bera2017aggressive,bera2016glmp}, there is only a small amount of work available on modeling dominance. Karg et al.~\cite{karg2010recognition} treat dominance as a dimension in the emotion space and identify emotions from gaits. Our approach to model dominance in virtual characters and to classify dominance from gaits combines data-driven techniques with a feature-based method inspired by these approaches.

\subsection{Behavior Modeling of Virtual Characters}
There is considerable literature in psychology, VR, and pedestrian simulation on modeling the behavior of pedestrians and virtual characters. Many rule-based methods have been proposed to model complex behaviors based on motor, perceptual, behavioral, and cognitive components~\cite{Terzopoulos}. There is extensive literature on modeling emergent behaviors, starting with Reynold's work~\cite{boids}.  Yeh et al.~\cite{Composite} describe velocity-based methods for modeling different behaviors, including aggression, social priority, authority, protection, and guidance. Other techniques use personality traits to model heterogeneous crowd behaviors~\cite{GuyPersonality,UPennOCEAN}. Perceptual or user studies are used to improve the behavior and rendering of virtual characters~\cite{randhavane2017f2fcrowds}. McDonnell et al.~\cite{eyeCrowds} use perceptual saliency to identify essential features that need to be varied to add visual variety to the appearance of avatars. \blue{Virtual character generation approaches have been developed based on the PAD model of emotion by Mehrabian et al.~\cite{mehrabian1996pleasure}. Vinayagamoorthy et al.~\cite{vinayagamoorthy2006building} summarize approaches that consider body posture and movement features in simulating emotions in virtual characters.} McHugh et al.~\cite{emotion} study the relationship between an agent's body posture and his or her perceived emotional state. Clavel et al.~\cite{clavel2009combining} combine facial and postural expressions to investigate the overall perception of basic emotions in virtual characters. Pelechaud et al.~\cite{pelachaud2009studies} use gestures to express different emotions in behaviors of virtual characters. Virtual agents with emotional capabilities have been used as museum guides~\cite{kopp2005conversational}. \blue{Su et al.~\cite{su2007personality} propose a rule-based method to model specific personality types. In this paper, we propose a data-driven method for generating virtual characters with a variety of dominance characteristics. Our approach is complementary to these methods and can be combined with them. We also provide a dominance classification algorithm that can be used to classify the dominance levels of the gaits generated using any method (e.g., Holden et al.~\cite{holden2017phase}).} \red{ In this paper, we aim to generate virtual characters with a variety of dominance characteristics.}

\section{Perception User Study}
\label{sec:perceptionStudy}
To be able to generate gaits that exhibit a range of dominance traits, we use a data-driven approach to compute a mapping between gaits and their dominance labels (Figure~\ref{fig:overview}). During precomputation, we use motion-captured gait datasets as input and obtain dominance labels for each gait using a perception user study. Using the results of this study, we establish a mapping between gaits and dominance. We use this mapping at runtime to generate gaits for virtual characters that have the desired dominance traits. In the rest of this section, we describe the details of this dominance perception study. 

\subsection{Study Goals}
The goal of this perception study was to obtain the dominance labels for gaits using three motion-capture datasets. 

\subsection{Gait Datasets}
\begin{figure}[t]
  \centering
  \includegraphics[width =\linewidth]{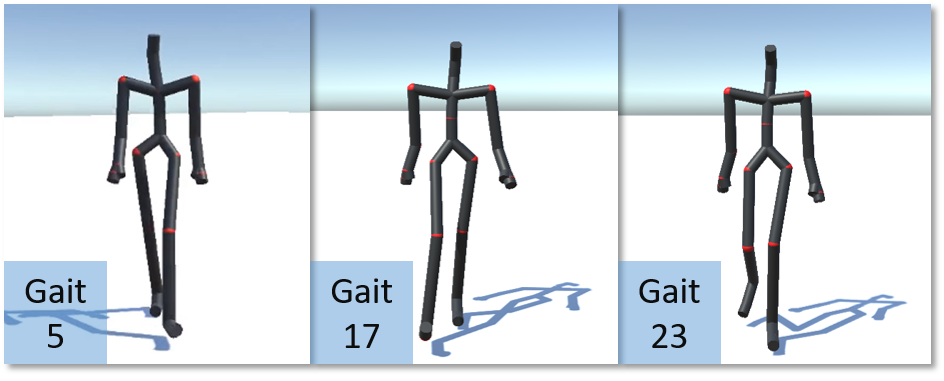}
  \caption{\textbf{Sample Visualization of the Gait Dataset:} We show sample visualizations of the gaits of $3$ individuals. Gait videos from $179$ motion-captured gaits were displayed to the participants in a web-based user study.}
  \label{fig:gaitVideo}
\end{figure}

We used three publicly available motion captured gait datasets: $36$ gaits from the CMU dataset~\cite{CMUGait}, $24$ gaits from the ICT dataset~\cite{narang2017motion}, and $119$ gaits from the BML dataset~\cite{ma2006motion}. Each gait was visualized using a skeleton mesh and rendered from the viewpoint of a camera looking at the mesh from the front  (Figure~\ref{fig:gaitVideo}). \blue{The visualizations were generated with the same frame rate with which they were captured.} The $179$ resulting videos were displayed to the participants in a web-based user study. 

\subsection{Participants} 
\blue{Since the data obtained from  Amazon Mechanical Turk has been shown to be at least as reliable as those obtained via traditional methods~\cite{buhrmester2011amazon}, we used it to recruit the participants. A total of} $703$ participant responses ($362$ female, $332$ male, $9$ preferred not to provide gender, $\bar{age} = 36.3$) were used to generate dominance labels.
For the smaller CMU and ICT gait datasets, each participant watched and rated a random subset of \blue{six} videos while other participants watched and rated a random subset of $12$ videos from the BML dataset. We assume that our gait dataset of $179$ videos is a representative sample of variations of dominance traits.

\subsection{Questionnaire}
We designed our questions using previous studies from psychology literature~\cite{rule2012perceptions, dunbar2005perceptions, ridgeway1987nonverbal, neave2003second}. We considered the adjectives submissive, non-confident, sluggish, withdrawn, non-aggressive, fearful, tense, dominant, confident, energetic, outgoing, aggressive, intimidating, and relaxed. These adjectives have been used in previous studies to assess the dominance traits of humans~\cite{rule2012perceptions, dunbar2005perceptions, ridgeway1987nonverbal, neave2003second}. Based on the results of a pilot user study, we decided to use a subset of these adjectives for the final study. For each video, participants were asked if they found the character to be \blue{\textit{submissive, withdrawn, dominant}, and \textit{confident}}. The participants answered each question on a \blue{five}-point Likert scale ranging from \blue{``strongly disagree" to ``strongly agree."} 

\subsection{Results}
\blue{For each gait $\textbf{G}_i$ in the dataset, we obtained a response $r^{adj}_{i, j}$ by a participant $j$ to an adjective $adj \in $ \{\textit{submissive, withdrawn, dominant, confident}\}. We analyzed the consistency in the participant responses using a method similar to that in Kleinsmith et al.~\cite{kleinsmith2013affective} to estimate how well the participants agreed.  We randomly divided the participant responses in two equal sets $\mathbb{P}_1$ and $\mathbb{P}_2$. For each adjective, we computed the average of the participant responses for each set $r^{adj, 1}_i$ and $r^{adj, 2}_i$:
\begin{eqnarray}
    r^{adj,1}_i &=& \frac{\sum_{j \in \mathbb{P}_1} r^{adj}_{i,j}}{n_{p,1}}, \\
    r^{adj,2}_i &=& \frac{\sum_{j \in \mathbb{P}_2} r^{adj}_{i,j}}{n_{p,2}},
\end{eqnarray}
where $n_{p,1}$ and $n_{p,2}$ are the cardinalities of $\mathbb{P}_1$ and $\mathbb{P}_2$, respectively. We computed the average error $e^{adj}$ between the two means $r^{adj, 1}_i$ and $r^{adj, 2}_i$:
\begin{eqnarray}
    e^{adj} &=& \frac{|r^{adj,1}_i - r^{adj,2}_i|}{N},
\end{eqnarray}
where $N=179$ is the number of gaits in the dataset. We observe an average error of $7.09\%$ between the two mean values for the four adjectives, indicating that the participant responses are reliable.}

\blue{In further analysis,} for each gait $\textbf{G}_i$ in the dataset, we calculated the mean of all participant responses ($r^{adj}_{i, j}$) to each adjective:
\begin{eqnarray}
    r^{adj}_i = \frac{\blue{\sum_{j=1}^{n_p}} r^{adj}_{i,j}}{n_p},
\end{eqnarray}
where $n_p$ is the number of participant responses collected and $adj$ is one of the four adjectives: \blue{\textit{submissive, withdrawn, dominant, confident}}.

\begin{figure*}[ht]
  \centering
  \includegraphics[width =\linewidth]{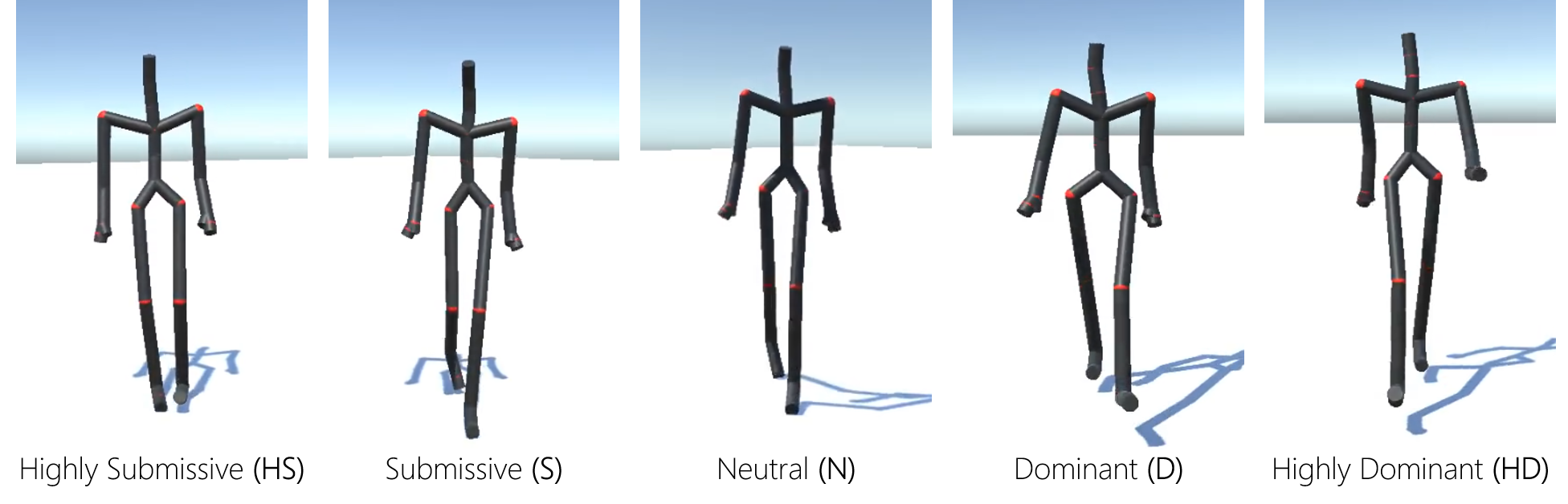}
  \vspace{-22pt}
  \caption{\textbf{Dominance spectrum:} Based on a \blue{perception user study}, we obtain dominance labels for motion-captured gaits. As an example, participants rated the character on the left as Highly Submissive (HS), whereas the character on the right as Highly Dominant (HD). According to the psychology literature, more leg and hand movement and erect posture is observed in a dominant gait as compared to a submissive gait. }
  \vspace{-10pt}
  \label{fig:dominance}
\end{figure*}

We \blue{also} analyzed the participant responses to the adjectives. Since all the adjectives capture general impressions of an individual's dominance, we expect a strong correlation between them~\cite{rule2012perceptions, dunbar2005perceptions, ridgeway1987nonverbal, neave2003second}. This is shown by the correlation coefficients in Table~\ref{tab:correl}.

\begin{table}[t]
\centering
\caption{\textbf{Correlation Between Dominance Adjectives}: We provide the correlation coefficients for the dominance adjectives. These adjectives are known to be closely related to an individual's dominance and are therefore highly correlated.}
\begin{tabular}{|c|c|c|c|c|}
\hline
           & Submissive & Withdrawn & Dominant & Confident \\ \hline
Submissive & 1.00       & 0.91      & -0.90    & -0.87     \\ \hline
Withdrawn  & 0.91       & 1.00      & -0.91    & -0.94     \\ \hline
Dominant   & -0.90      & -0.91     & 1.00     & 0.94      \\ \hline
Confident  & -0.87      & -0.94     & 0.94     & 1.00      \\ \hline
\end{tabular}
\vspace{-15pt}
\label{tab:correl}
\end{table}

\subsection{Data-Driven Dominance Mapping}
The high correlations between the adjectives suggest that the underlying dominance factor in the participant responses can be obtained using factor analysis methods. Therefore, we performed Principal Component Analysis (PCA) on the average participant responses for each video revealing that a single factor can account for $93.78\%$ variance. We use this factor to combine the responses $r^{sub}_{i}$, $r^{with}_{i}$, $r^{dom}_{i}$, $r^{conf}_{i}$ for a gait $\textbf{G}_i$ into a scalar value $r_{i} \in \mathbb{R}$:

\begin{eqnarray}
    r_{i} =  0.43*r^{dom}_{i} + 0.54*r^{conf}_{i} -0.44*r^{sub}_{i} -0.57*r^{with}_{i}. \label{eq:domLabel}
\end{eqnarray}
The negative coefficients of responses to the submissive and withdrawn adjectives correspond to the fact that their meanings are opposite from those of the dominant and confident adjectives. We also normalize the values such that $r_{i} \in [-1, 1]$ \blue{with $-1$ denoting the minimum observed value and $1$ denoting the maximum observed value}.

Since there can be a disagreement between observers about perceived dominance~\cite{kleinsmith2013affective}, instead of using a scalar value for the  dominance of a gait, we use \blue{five} classes as dominance labels: \textit{(Highly Submissive (\textbf{HS}), Submissive (\textbf{S}), Neutral (\textbf{N}), Dominant (\textbf{D}), Highly Dominant (\textbf{HD}))}.

According to McCrae et al.~\cite{mccrae1992introduction}, most people lie somewhere in the middle of the personality scale. Using the scalar values of dominance $r_i$, we obtain the dominance label $d_{i}$ for a gait $\textbf{G}_i$ by dividing the gaits into \blue{five} dominance levels with Neutral level ($N$) containing most gaits:
\begin{equation}
    d_{i} = 
\begin{cases}
   \textbf{HS}\, & \text{if } -1 \leq r_i < -0.8\\
   \textbf{S}\, & \text{if } -0.8 \leq r_i < -0.5\\
   \textbf{N}\, & \text{if } -0.5 \leq r_i \leq 0.5\\
   \textbf{D}\, & \text{if } 0.5 < r_i \leq 0.8\\
   \textbf{HD}\, & \text{if } 0.8 < r_i \leq 1\\
\end{cases} \label{eq:domLabels}
\end{equation}

\begin{figure}[ht]
    \includegraphics[width =1\linewidth]{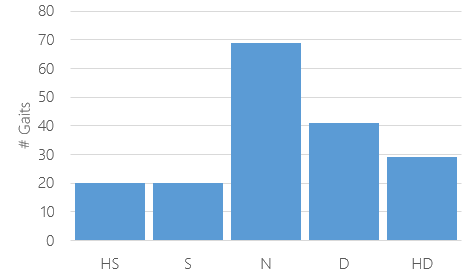}
    \vspace{-22pt}
    \caption{\textbf{Dataset Variation:} We divide $179$ gaits into $5$ dominance levels using a \blue{perception user study}: \textit{(Highly Submissive (\textbf{HS}), Submissive (\textbf{S}), Neutral (\textbf{N}), Dominant (\textbf{D}), Highly Dominant (\textbf{HD}))}.}
    \vspace{-15pt}
    \label{fig:datasetVariation}
\end{figure}

Figure~\ref{fig:dominance} shows sample gaits and their computed labels. Figure~\ref{fig:datasetVariation} shows the variation in gait dominance in accordance with our assumption that the gaits from the three datasets capture the variation in dominance traits.

\subsection{Interactive Virtual Character Gait Generation} \label{sec:charGeneration}
\begin{figure}
    \centering
    \includegraphics[height=150pt]{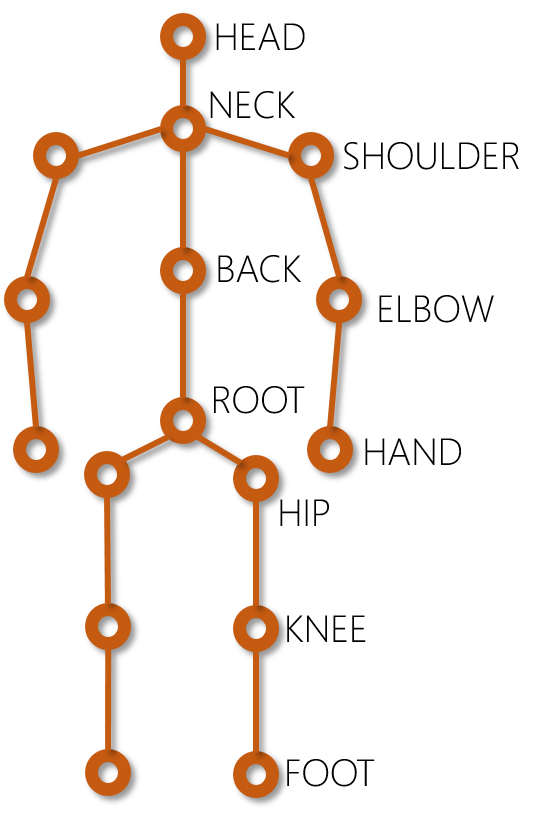}
    \vspace{-10pt}
    \caption{\textbf{Character Representation}: We represent a character as a skeleton with $16$  body joints to compute different gait features. The overall configuration of a skeleton, which is used to extract the features, is defined using these joint features.}
    \vspace{-15pt}
    \label{fig:Joints}
\end{figure}

\red{\textbf{Notation:}
We represent a character as a skeleton $\textbf{S}$ with a mesh attached to it. A skeleton is represented as a set of joints $\textbf{S} = \{J_1, J_2, J_3, ..., J_{n}\}$. The number of joints $n$ depends on the methods used to generate gaits (e.g., motion capture markers or the motion blending algorithm). For this work, we assume a skeleton with $16$ joints (Figure\mbox{~\ref{fig:Joints}}). A pose $P \in \mathbb{R}^{48}$ of a character is a set of 3D positions of each joint $J_i$. A gait $\textbf{G}_i$ is a set of 3D poses ${P_1, P_2,..., P_{\tau}}$ where $\tau$ is the number of frames in the input gait. For navigation in the virtual environment, we represent the state of the virtual environment by $\textbf{E}$ that includes the positions and velocities of all virtual characters in the environment.}

At runtime, our character generation algorithm (Algorithm~\ref{Algorithm}) takes the desired dominance level $d_{des}$ as input. \red{and chooses a gait $G_{des}$ from the labeled dataset associated with the desired dominance level.} \blue{We can also specify any other movement control requirements. For example, some navigation algorithms constrain a character's walking speed to a maximum value~\cite{van2011reciprocal}. We represent these requirements as a boolean function $f(G_i)$ that takes a gait $G_i$ as input and returns true if $G_i$ satisfies the requirement and false otherwise. We find the set of gaits $\mathbb{G}$ from the labeled dataset associated with the desired dominance level:}
\begin{eqnarray}
    \blue{\mathbb{G} = G_i \mid d_{des} = d_{i} \text{ and }f(G_i) = true.}
\end{eqnarray}

\begin{table}
\centering
\caption{\blue{\textbf{Average Frame Update Time}: We present the average frame update time for generating gaits of virtual characters with different dominance traits. We compare the performance to an algorithm that does not consider any dominance traits.}}
\label{tab:timing}
{\blue{\begin{tabular}{|c|c|c|}
\hline
Number of Characters & \begin{tabular}[c]{@{}c@{}}Without Dominance \\ (in ms)\end{tabular} & \begin{tabular}[c]{@{}c@{}}With Dominance\\ (in ms)\end{tabular} \\ \hline
1 & 11.52 & 11.53 \\ \hline
2 & 11.53 & 11.52 \\ \hline
5 & 11.80 & 11.53 \\ \hline
10 & 11.50 & 11.85 \\ \hline
20 & 11.88 & 11.79 \\ \hline
50 & 11.55 & 11.81 \\ \hline
100 & 12.38 & 12.59 \\ \hline
\end{tabular}}}
\end{table}

\blue{We choose a gait $G_{des} \in \mathbb{G}$ using random selection and} update the joint positions of the character in the virtual world using the joint positions from $G_{des}$. After updating the joint positions of the character, its root position can be calculated using any navigation algorithm~\cite{RVO} to generate the character's movement in the virtual world. 

\begin{algorithm}[t]
\KwIn{desired dominance level $d_{des}$, gaits $\textbf{G}_i, i \in \{1, ..., 179\}$ and their dominance labels $d_i, i \in \{1, ..., 179\}$, time $t$}
\KwOut{character's joint positions for joints $J_{k}, k \in \{1, ..., 16\}$}
\nl $\textbf{G}_{des}$ = $\textbf{G}_i$ where $d_i = d_{des}$\\
\nl $P_{des}$ = $P_t$ where $P_t \in \textbf{G}_{des}$ \\
\nl \For{joint $J_{k}$ in skeleton S}{
    Position $J_k =$ Position $ J_k^{des}$ where $J_k^{des} \in P_{des}$
    }
\nl Root position $J_{1} = navigation(\textbf{E}, t-1)$
\caption{{\bf Interactive Virtual Character Generation \label{Algorithm}}}
\end{algorithm}

\subsubsection{\blue{Interactivity Results}}
\blue{Using our Virtual Character Gait Generation method, we can generate gaits for virtual characters at interactive rates. We present the average frame update time in Table~\ref{tab:timing}. The results show that our algorithm can generate gaits for tens of virtual characters at interactive rates.}


\begin{table}
\centering
\caption{\textbf{Gait Features}: Based on visual perception and psychology literature, we extract these features from an input gait. Since both posture and movement features are essential for an accurate prediction of an individual's affective state~\cite{kleinsmith2013affective}, we define features that include both the posture and the movement features.}
\begin{tabular}{|l|l|c|}
\hline
\multicolumn{1}{|c|}{Type}                                                         & \multicolumn{1}{c|}{Description}                                                            & Category                   \\ \hline
Volume                                                                             & Bounding box                                                                                & \multirow{13}{*}{Posture}  \\ \cline{1-2}
\multirow{5}{*}{Angle}                                                             & At neck by shoulders                                                                        &                            \\ \cline{2-2}
                                                                                   & \begin{tabular}[c]{@{}l@{}}At right shoulder by \\ neck and left shoulder\end{tabular}      &                            \\ \cline{2-2}
                                                                                   & \begin{tabular}[c]{@{}l@{}}At left shoulder by \\ neck and right shoulder\end{tabular}      &                            \\ \cline{2-2}
                                                                                   & At neck by vertical and back                                                                &                            \\ \cline{2-2}
                                                                                   & At neck by head and back                                                                    &                            \\ \cline{1-2}
\multirow{5}{*}{Distance}                                                          & \begin{tabular}[c]{@{}l@{}}Between right hand \\ and the root joint\end{tabular}            &                            \\ \cline{2-2}
                                                                                   & \begin{tabular}[c]{@{}l@{}}Between left hand \\ and the root joint\end{tabular}             &                            \\ \cline{2-2}
                                                                                   & \begin{tabular}[c]{@{}l@{}}Between right foot \\ and the root joint\end{tabular}            &                            \\ \cline{2-2}
                                                                                   & \begin{tabular}[c]{@{}l@{}}Between left foot \\ and the root joint\end{tabular}             &                            \\ \cline{2-2}
                                                                                   & \begin{tabular}[c]{@{}l@{}}Between consecutive\\  foot strikes (stride length)\end{tabular} &                            \\ \cline{1-2}
\multirow{2}{*}{Area}                                                              & \begin{tabular}[c]{@{}l@{}}Triangle between \\ hands and neck\end{tabular}                  &                            \\ \cline{2-2}
                                                                                   & \begin{tabular}[c]{@{}l@{}}Triangle between \\ feet and the root joint\end{tabular}         &                            \\ \hline
\multirow{5}{*}{Speed}                                                             & Right hand                                                                                  & \multirow{16}{*}{Movement} \\ \cline{2-2}
                                                                                   & Left hand                                                                                   &                            \\ \cline{2-2}
                                                                                   & \blue{Head}                                                                                        &                            \\ \cline{2-2}
                                                                                   & Right foot                                                                                  &                            \\ \cline{2-2}
                                                                                   & Left foot                                                                                   &                            \\ \cline{1-2}
\multirow{5}{*}{\begin{tabular}[c]{@{}l@{}}Acceleration \\ Magnitude\end{tabular}} & Right hand                                                                                  &                            \\ \cline{2-2}
                                                                                   & Left hand                                                                                   &                            \\ \cline{2-2}
                                                                                   & \blue{Head}                                                                                        &                            \\ \cline{2-2}
                                                                                   & Right foot                                                                                  &                            \\ \cline{2-2}
                                                                                   & Left foot                                                                                   &                            \\ \cline{1-2}
\multirow{5}{*}{\begin{tabular}[c]{@{}l@{}}Movement \\ Jerk\end{tabular}}          & Right hand                                                                                  &                            \\ \cline{2-2}
                                                                                   & Left hand                                                                                   &                            \\ \cline{2-2}
                                                                                   & \blue{Head}                                                                                        &                            \\ \cline{2-2}
                                                                                   & Right foot                                                                                  &                            \\ \cline{2-2}
                                                                                   & Left foot                                                                                   &                            \\ \cline{1-2}
Time                                                                               & One gait cycle                                                                              &                            \\ \hline
\end{tabular}
\label{tab:features}
\end{table}
\section{Gait Dominance Classification}
During runtime simulation, new gaits can be generated using motion blending techniques~\cite{feng2012analysis}. To predict the dominance traits of new gaits, we use a feature-based approach. After the perception user study, we get a dominance label (Equation~\ref{eq:domLabels}) for each gait in the motion-captured gait dataset.  If we also extract the feature values from each gait in the motion-captured dataset, then we can train a classifier using the annotated data. This classifier can then classify the dominance traits of any new input gait. For this purpose, it is necessary to understand the features that cause a gait to be perceived as dominant or submissive. We describe these features below. 

\subsection{Feature Extraction}
Previous studies in psychology literature~\cite{furley2012nonverbal, weisfeld1982erectness,curran2016methods,richards1991perceptions,eisenburg1937expressive} have shown that factors like postural erectness, postural openness, slumped shoulders, walking speed, etc. affect the perception of dominance. Previous work on recognition of emotions and affective states from gaits has also determined features in the form of joint angles, distances, and velocities of the joints, and space occupied by the body~\cite{crenn2016body,kleinsmith2013affective}. Since both posture and movement features are essential for an accurate prediction of an individual's affective state~\cite{kleinsmith2013affective}, we define features that include both the posture and the movement features. We list these gait features in Table~\ref{tab:features} and describe them in detail below.

We represent the extracted features of a gait $\textbf{G}_i$ as a vector $F_i \in \mathbb{R}^{29}$. Given the gait features $F_i$, we represent the predicted dominance label as $d_i^{pred}$.

\subsubsection{Posture Features}
\noindent \red{\textbf{Openness}:} Openness of the limbs has been shown to affect dominance traits. Exhibiting slightly spread hands and legs is perceived as dominant, whereas minimizing the occupied space by pulling hands and legs in towards the torso is perceived as submissive~\cite{furley2012nonverbal,eisenburg1937expressive,weisfeld1982erectness,hall2005nonverbal}. We model this as the volume of the bounding box, the area of the triangle defined by both hands and neck, the area of the triangle defined by both feet and root, the three angles induced by the triangle formed by both shoulders and neck, and the distances between hands and feet and the root joint. We also use stride length as a feature to measure openness.

\noindent \red{\textbf{Erectness}:} Dominant gaits have been observed to involve an erect posture~\cite{furley2012nonverbal,eisenburg1937expressive,weisfeld1982erectness,hall2005nonverbal}. In contrast, a submissive gait involves a slouched posture with the head and chin pointing down. We represent the head orientation as the angle formed by the head and back joint at the neck. A slouched or erect posture is represented by the angle between the vertical and back formed at the neck.

\subsubsection{Movement Features}
\noindent \red{\textbf{Joint Movements}:} Movement of body parts such as hand and leg joints are perceived as dominant, whereas less joint movement is perceived as submissive~\cite{furley2012nonverbal,eisenburg1937expressive,weisfeld1982erectness,hall2005nonverbal}. We model this based on the magnitude of velocity, acceleration, and movement jerk (derivative of acceleration) of the hands, feet, and head.

\noindent \red{\textbf{Gait Speed}:} Fast-moving people are perceived as more dominant than slow-moving people. Low walking speeds are also regarded as less confident, and high walking speeds are perceived as more confident~\cite{clark1986efficiency}. We model this feature by the time taken to complete one gait cycle and the stride length.

\subsubsection{\blue{Feature Representation}}
Given an input gait $\textbf{G}_i$, we compute features $F_{i,j}$ for each pose $P_j$ corresponding to a frame in the gait. We define the gait feature $F_i$ as the average of $F_{i,j}, j=\{1,2,..,\tau\}$:

\begin{eqnarray}
    F_i = \frac{\sum_{i} F_{i,j}}{\tau},
\end{eqnarray}

We also append gait time and stride length features to $F_i$.

\subsection{Dominance Classification}
Given the feature values extracted using the above method and the dominance labels obtained from the perception user study, we train a dominance classifier. For each gait $\textbf{G}_i$ in the gait dataset, we have a vector of feature values $F_i$ and a dominance label $d_i$. Before training the classifier, we normalize the feature values to $[-1, 1]$ \blue{with $-1$ denoting the minimum value of the feature and $1$ denoting the maximum value of the feature}. We use Support Vector Machines (SVMs) to classify the features similar to previous approaches for emotion identification from gait features~\cite{crenn2016body}. We use an RBF kernel with a  one-versus-rest decision function of shape\blue{~\cite{crenn2016body,weston1999support}}.



\section{Validation User Study}
\label{sec:validationStudy}
We performed a user study to validate our approach (Algorithm~\ref{Algorithm}) for generating virtual characters with desired dominance levels using our mapping. In this section, we give the details of this user study.

\begin{figure*}[t]
    \centering
    \includegraphics[width=1.0\linewidth]{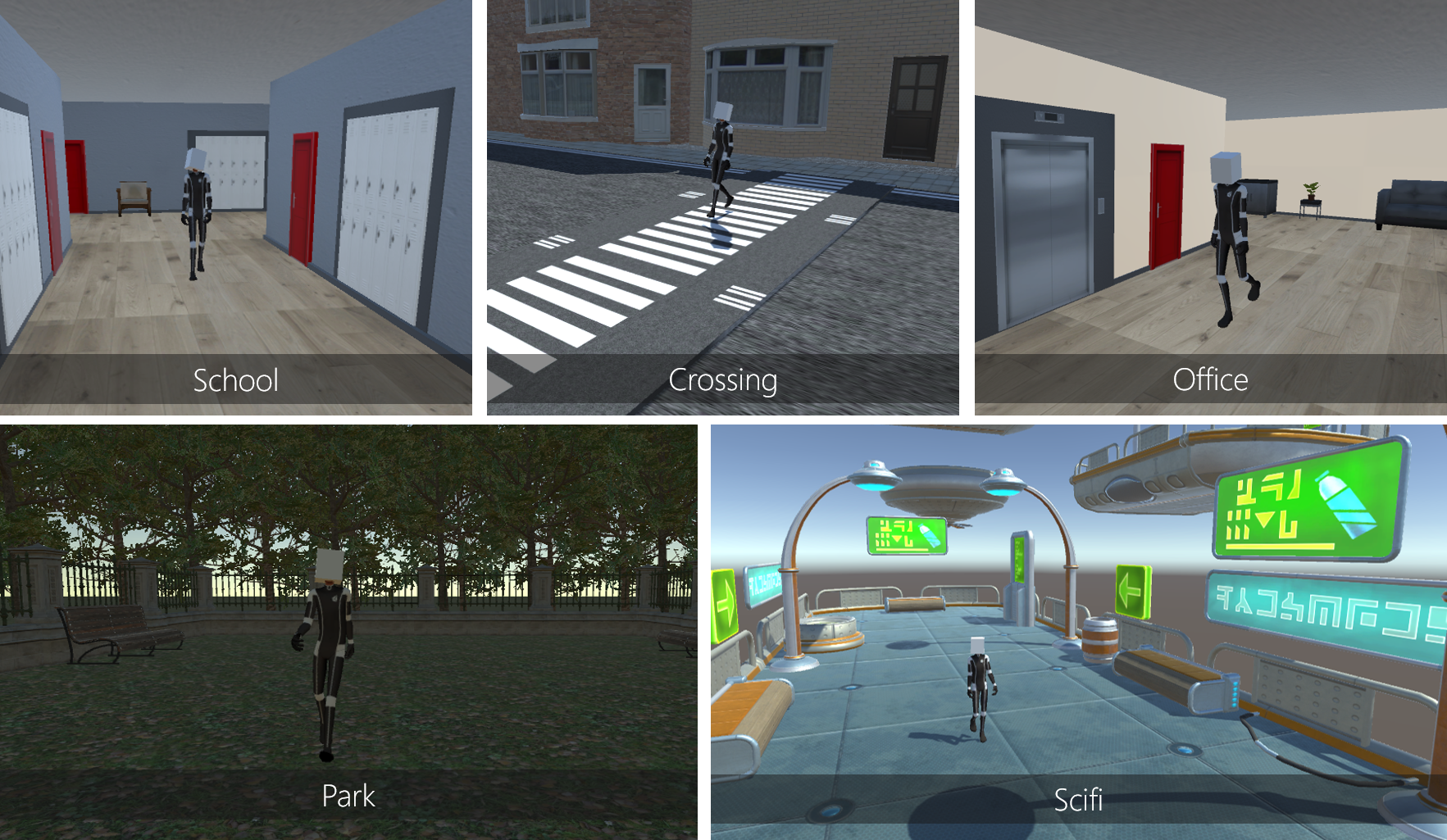}
    \vspace{-15pt}
    \caption{\textbf{Scenarios:} Our \blue{validation user study} included $5$ scenarios, including indoor, outdoor, residential, and fantastical scenes.}
    \vspace{-15pt}
    \label{fig:scenarios}
\end{figure*}

\subsection{Study Goals and Expectation}
This study aimed to show that virtual characters generated using our approach (Algorithm~\ref{Algorithm}) could exhibit a variety of submissive and dominant personality traits. In particular, we propose the following hypothesis:

\textbf{Hypothesis}: Our data-driven approach can be used to generate virtual characters with varying levels of dominance.

\subsection{Experimental Design}
The study was conducted based on a within-subjects design. Participants were shown \blue{five} scenes with a virtual character walking in different environments. Participants performed $10$ trials per scene, corresponding to $10$ virtual characters with varying levels of predicted dominance. The order of the scenes and the dominance levels of the virtual characters were counterbalanced. Participants performed the study using HTC Vive HMD. Participants could look around in the virtual environment by rotating their heads and could also walk in the tracking space, but there was no interaction with the virtual characters (Figure \ref{fig:frontCover} (top)). 

\subsubsection{Procedure}
After welcoming the participants, they were instructed on the overall process and purpose of the study. They were informed that the study involved using immersive hardware (HMD) and that it may cause nausea and slight discomfort. The experiment was approved by Institutional Review Boards and the Office of Human Research Ethics. Before beginning the experiment, the participants were invited to read and agree to the approved consent form. The participants were asked to provide optional demographic information about their gender and age. The study required approximately $30$ minutes, and participants were compensated with a gift card worth \$$5$.

\subsubsection{Participants}
We recruited $51$ participants ($38$ male, $13$ female,  $\bar{x}_{age}$ = $23.2$) from the staff and students of a university. 

\subsubsection{Scenarios}
We evaluated the dominance characteristics of the virtual characters in \blue{five} different scenarios (Figure~\ref{fig:scenarios}):

\begin{itemize}
    \item \textbf{Crossing}: This outdoor scenario had a virtual character crossing a street in a residential environment.
    \item \textbf{Office}: This indoor scenario had a virtual character walking in an office environment.
    \item \textbf{Park}: This outdoor scenario consisted of a virtual character walking in a park.
    \item \textbf{School}: This indoor scenario consisted of a virtual character walking in a school corridor.
    \item \textbf{Scifi}: This fantastical scenario involved a virtual character walking on a floating platform.
\end{itemize}

In each of the scenarios, we used a template character. The character's clothing and appearance were chosen to minimize the dominance cues from appearance and were kept constant through-out the experiment. The character's face was also hidden behind a solid cube to avoid facial cues. We generated the character's gait with \blue{two} gaits from each dominance level, resulting in a total of $10$ trials. From now on, we represent these gaits as: $HS1\,, HS2\,, S1\,, S2\,, N1\,, N2\,, D1\,, D2\,, HD1\,, HD2$.



\subsubsection{Questions} 
As in the \blue{perception user study} (Section~\ref{sec:perceptionStudy}), we asked the participants whether they found the character to be \blue{\textit{submissive, withdrawn, dominant}, and \textit{confident}} on a \blue{five}-point Likert scale ranging from \blue{``strongly disagree" to ``strongly agree."} 

\subsection{Discussion}
Here we present and analyze the participant responses. \blue{For each scene, we obtained the participant responses $r^{sub}_{i,j}$, $r^{with}_{i,j}$, $r^{dom}_{i,j}$, $r^{conf}_{i,j}$ corresponding to a character simulated with a gait $G_i$, participant $j$, and the four adjectives, \textit{submissive, withdrawn, dominant}, and \textit{confident}. For each participant, we converted the responses to the four adjectives into a scalar value $r_{i,j}$ using the principal component in Equation~\ref{eq:domLabel}}:
\begin{eqnarray}
    \blue{r_{i,j} =  0.43*r^{dom}_{i,j} + 0.54*r^{conf}_{i,j} -0.44*r^{sub}_{i,j} -0.57*r^{with}_{i,j}.}
\end{eqnarray}

\blue{For each scene, we computed the mean of the participant responses for each of the $10$ gaits:}
\begin{eqnarray}
   \blue{ r_{i} = \frac{\sum_{j=1}^{n_p} r_{i,j}}{n_p}},
\end{eqnarray}
\blue{where $n_p$ is the number of participant responses collected for the character simulated with the gait $G_i$.} We present these mean participant responses in Table~\ref{tab:means}. We normalized the means $\in [-1, 1]$, where a higher value indicates higher dominance. We observed that gaits from higher dominance levels have higher mean values, as predicted by our algorithm. 

\begin{table}
\centering
\caption{\textbf{Reported Dominance:} We present the mean values of the normalized participant responses $\in [-1, 1]$. We generated $10$ characters for each scenario using \blue{two} gaits from each dominance level: $HS, S, N, D, HD$. Participants reported higher dominance for more dominant gaits, as predicted by our algorithm, across all the scenarios.}
\begin{tabular}{|c|c|c|c|c|c|}
\hline
 & Crossing & Office & Park & School & Scifi \\ \hline \rowcolor[HTML]{C0C0C0} 
HS1 & -0.88 & -0.90 & -0.94 & -0.93 & -0.97 \\ \hline \rowcolor[HTML]{C0C0C0} 
HS2 & -0.82 & \blue{\textbf{-0.41}} & -0.82 & -0.80 & -0.80 \\ \hline 
S1 & \blue{\textbf{-0.45}} & -0.53 & -0.66 & -0.62 & -0.55 \\ \hline
S2 & -0.59 & \blue{\textbf{-0.45}} & -0.58 & \blue{\textbf{-0.48}} & -0.55 \\ \hline \rowcolor[HTML]{C0C0C0} 
N1 & 0.17 & 0.05 & -0.10 & -0.09 & -0.06 \\ \hline \rowcolor[HTML]{C0C0C0} 
N2 & -0.10 & 0.07 & -0.27 & -0.06 & -0.42 \\ \hline
D1 & 0.59 & 0.62 & \blue{\textbf{0.48}} & \blue{\textbf{0.45}} & 0.52 \\ \hline
D2 & 0.79 & 0.72 & 0.76 & 0.78 & 0.70 \\ \hline \rowcolor[HTML]{C0C0C0} 
HD1 & 1.00 & 0.84 & 0.97 & 0.80 & 0.90 \\ \hline \rowcolor[HTML]{C0C0C0} 
HD2 & 0.95 & 0.99 & 0.97 & 0.98 & 0.91 \\ \hline
\end{tabular}
\vspace{2pt}
\label{tab:means}
\end{table}

We also compare the means of participant responses for pairs of gaits with different levels of dominance using paired samples t-tests. We present the p-values from gaits with consecutive levels of predicted dominance in Table~\ref{tab:pvalues}. \blue{In the realistic outdoor \textit{Park} scenario, we observed significant differences ($p < 0.05$) in all the comparisons. For the other outdoor scenario, \textit{Crossing},} the indoor scenarios, \textit{Office} and \textit{School}, and the fantastical scenario, \textit{Scifi}, most comparisons show statistically significant differences. These results support our hypothesis that our data-driven approach (Algorithm~\ref{Algorithm}) can be used to generate virtual characters with varying levels of dominance.

\begin{table}
\caption{\textbf{Comparison of Means:} We compare the means of participant responses for pairs of gaits with different levels of dominance using paired samples t-tests. This table presents the p-values for these comparisons \blue{(highlighted values indicate $p \geq 0.05$)}. We observe statistically significant differences ($p < 0.05$) in the means of reported dominance for pairs of gaits with different predicted dominance levels. These results support our hypothesis that our data-driven approach can be used to generate virtual characters with varying levels of dominance.}
\centering
\begin{tabular}{|c|c|c|c|c|c|}
\hline
Comparison & Crossing & Office & Park & School & Scifi \\ \hline
HS1, S1 & 0.00 & 0.00 & 0.00 & 0.00 & 0.00 \\ \hline
HS1, S2 & 0.00 & 0.00 & 0.00 & 0.00 & 0.00 \\ \hline
HS2, S1 & 0.00 & \blue{\textbf{0.21}} & 0.04 & 0.04 & 0.01 \\ \hline
HS2, S2 & 0.03 & \blue{\textbf{0.64}} & 0.00 & 0.00 & 0.03 \\ \hline
S1, N1 & 0.00 & 0.00 & 0.00 & 0.00 & 0.00 \\ \hline
S1, N2 & 0.00 & 0.00 & 0.00 & 0.00 & \blue{\textbf{0.29}} \\ \hline
S2, N1 & 0.00 & 0.00 & 0.00 & 0.00 & 0.00 \\ \hline
S2, N2 & 0.00 & 0.00 & 0.01 & 0.00 & \blue{\textbf{0.27}} \\ \hline
N1, D1 & 0.00 & 0.00 & 0.00 & 0.00 & 0.00 \\ \hline
N1, D2 & 0.00 & 0.00 & 0.00 & 0.00 & 0.00 \\ \hline
N2, D1 & 0.00 & 0.00 & 0.00 & 0.00 & 0.00 \\ \hline
N2, D2 & 0.00 & 0.00 & 0.00 & 0.00 & 0.00 \\ \hline
D1, HD1 & 0.00 & 0.02 & 0.00 & 0.00 & 0.00 \\ \hline
D1, HD2 & 0.00 & 0.00 & 0.00 & 0.00 & 0.00 \\ \hline
D2, HD1 & 0.02 & \blue{\textbf{0.16}} & 0.02 & \blue{\textbf{0.84}} & 0.03 \\ \hline
D2, HD2 & \blue{\textbf{0.05}} & 0.00 & 0.01 & 0.01 & 0.01 \\ \hline
\end{tabular}
\label{tab:pvalues}
\end{table}

\begin{figure}[t]
    \centering
    \includegraphics[width=1.0\linewidth]{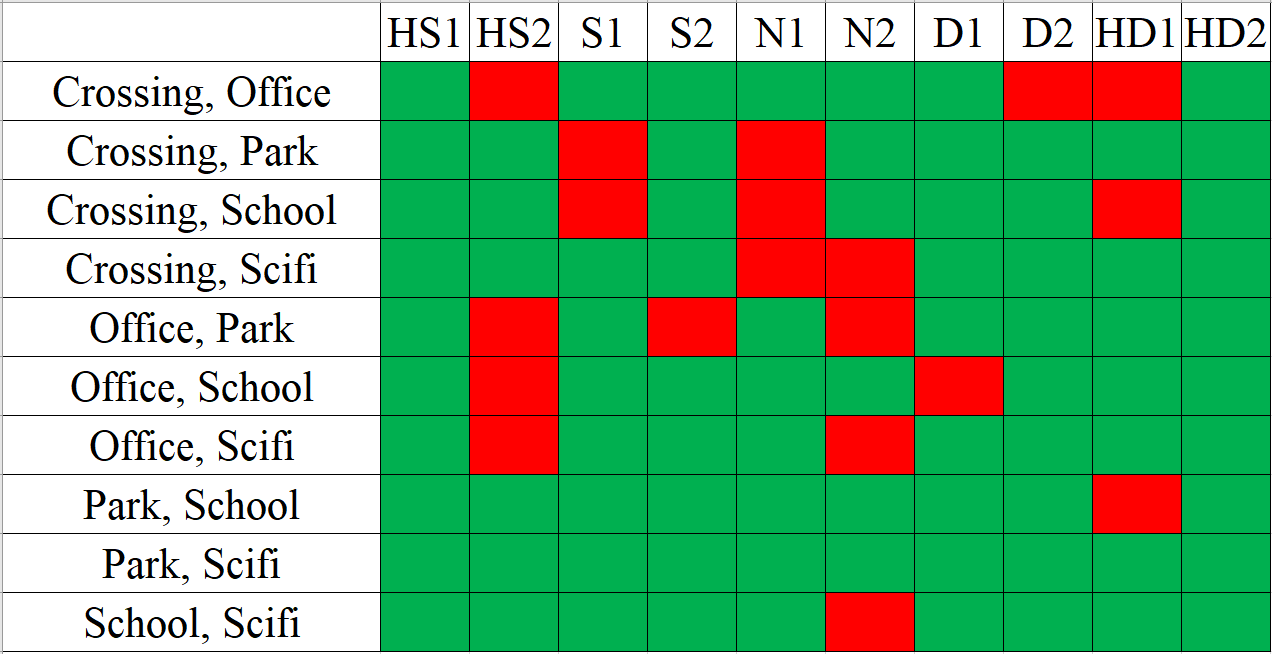}
    \vspace{-15pt}
    \caption{\blue{\textbf{Consistency Across Scenarios}: We perform paired samples t-tests between scenarios to assess whether the participant responses for a gait remain consistent across a variety of scenarios. We present a visualization of the p-values obtained for these comparisons. We color the cells where we observed a significant difference $(p < 0.05)$ between mean participant responses red and we color the cells where we did not observe a significant difference $(p > 0.05)$ green. For most of the gaits, there was not a significant difference $(p > 0.05)$ between mean participant responses across scenarios, indicating consistent dominance levels irrespective of the scenario.}}
    \vspace{-5pt}
    \label{fig:consistency}
\end{figure}

We conducted the user study for \blue{five} different scenarios. We performed paired sample t-tests for pairs of scenarios to assess whether the dominance levels of the generated characters varied significantly across different scenarios. For most of the characters, the dominance levels did not exhibit significant differences across scenarios, as indicated in Table~\ref{fig:consistency}.

\section{\blue{Dominance Classification Accuracy}}
In this section, we present the results of the gait dominance classification algorithm. We divide the datasets into training and testing datasets. For each gait $\textbf{G}_i$ in the dataset, we have a dominance label $d_i$ computed from the perception user study. We train our gait dominance classification algorithm using the training dataset and predict the dominance $d_i^{pred}$ for each gait $\textbf{G}_i$ in the testing dataset. We define the accuracy of our algorithm as the percentage of correctly predicted gait dominance levels from the testing dataset. Here, we treat dominance labels $d_i$ as the ground truth.

Table~\ref{tab:accuracy} presents the results of our experiments. We considered two sets of dominance levels:
\begin{itemize}
    \item \blue{Five} dominance levels: $HS\,,S\,,N\,,D\,,HD$ (Equation~\ref{eq:domLabels}) and
    \item \blue{Three} dominance levels: $S\,,N\,,D$ where $S = HS \cup S$ and  \\
    $D = D \cup HD$
\end{itemize}

\begin{table}
\caption{\textbf{Accuracy of Gait Dominance Classification}: We present the percentage accuracy of our gait dominance classification algorithm. We trained our algorithm on the training datasets and tested it on the testing datasets. For the datasets where the training and testing datasets were the same, we performed a $10$-fold cross-validation with $2000$ iterations.}
\centering
\begin{tabular}{|c|c|c|c|}
\hline
\begin{tabular}[c]{@{}c@{}}Training\\   Dataset\end{tabular} & \begin{tabular}[c]{@{}c@{}}Testing \\ Dataset\end{tabular} & 3 Levels & 5 Levels \\ \hline
All & All & 72.94 & 61.33 \\ \hline
BML + CMU & BML + CMU & 74.02 & 63.24 \\ \hline
BML + ICT & BML + ICT & 68.56 & 48.31 \\ \hline
BML & BML & 71.05 & 50.41 \\ \hline
BML & CMU & 77.78 & 83.33 \\ \hline
BML & ICT & 62.50 & 45.83 \\ \hline
BML & CMU + ICT & 73.33 & 71.67 \\ \hline
\end{tabular}
\label{tab:accuracy}
\end{table}

We also considered different training and testing datasets. We trained our algorithm using all the three datasets, using BML and CMU/ICT, and using BML only. We did not train on only CMU and ICT because these datasets contain very few samples. For the datasets where training and testing datasets were the same, we performed a $10$-fold cross-validation with $2000$ iterations. We observed consistent accuracy for most of the datasets. The performance deteriorated for the ICT dataset. A possible reason for this deterioration is that the dataset is noisier than the BML and CMU datasets. We observe an average accuracy of \texttildelow 73\% and \texttildelow 61\% when using three dominance levels and five dominance levels for classification, respectively. 
\section{Application}
\label{sec:applicationStudy}
In this section, we present an application of our approach that models interpersonal social relations between virtual characters. Among the many dimensions of interpersonal social relations, we consider the ``vertical" dimension, which relates to power, dominance, status, and social hierarchy~\cite{hall2005nonverbal}. Vertical dimension constructs are defined to be included in situations where there is
\begin{itemize}
    \item a situationally defined power, expertise, or status (for example, a teacher-student relationship),
    \item a self-reported dominant or assertive personality, or
    \item a perceived and/or rated impression of dominance.
\end{itemize}

\begin{figure}[t]
  \centering
  \includegraphics[width =\linewidth]{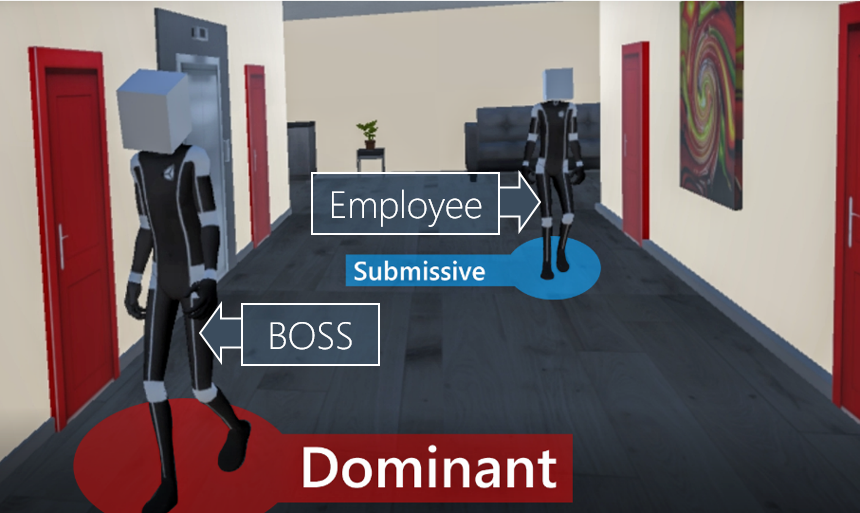}
  \vspace{-15pt}
  \caption{\textbf{Interpersonal Social Relationship Between Virtual Characters}: Our approach can be used to realize the vertical dimension of interpersonal social relationships. Members of a pair of dominant and submissive characters generated using our method were perceived by participants as being a boss or an employee depending on their dominance level in our \blue{application user study}.}
  \label{fig:application}
\end{figure}

Our approach to generating virtual characters with different perceived dominance traits can be used to simulate situations where there is a relationship belonging to the vertical dimension between virtual characters. To support this argument, we created simplified scenarios where the vertical dimension of interpersonal social relationships between virtual characters was realized and validated with a user study (Figure~\ref{fig:application}). We discuss the user study in this section.

\subsection{Study Goals}
We propose the following simplified hypothesis:

\noindent \textbf{Hypothesis:} Pairs of characters with different submissive and dominant characteristics can be used to realize the vertical dimension of interpersonal social relations.

\subsection{Experimental Design}
As in the \blue{validation user study} (Section~\ref{sec:validationStudy}), we used a within-subjects design in which the participants were asked to participate in two scenarios. 

\subsubsection{Procedure}
The procedure was like that in the \blue{validation user study}. Participants performed four trials of each of the two scenarios and were compensated with a gift card worth \$5.

\subsubsection{Participants}
We recruited $20$ participants ($15$ male, $5$ female,  $\bar{x}_{age}$ = $26.2$) from the staff and students of a university.

\subsubsection{Scenarios}
We wanted to evaluate whether the {\em vertical dimension}, as observed in real-world situations, can be realized in VR by generating virtual characters with varying dominance traits using our approach. We used two scenarios from the \blue{validation user study} in this experiment: \textit{Office} and \textit{School}. In each of the scenarios, two characters were generated with varying levels of dominance. Each participant performed four trials of each scenario in a randomized order with the two characters having dominance levels as described in Table~\ref{tab:userStudyTrials}. A single \textit{HS} gait and a single \textit{HD} gait was used for the trials.

We assumed that, if our virtual character generation approach can generate dominant and submissive characters, then the vertical dimension of the relationship between a teacher and a student will cause participants to choose the dominant character as the teacher and the submissive character as the student. Similarly, our assumption in the \textit{Office} case was that because of the dominance relationship between a boss and an employee, participants will choose the dominant character as the boss and the submissive character as the employee. 

\begin{table}
\centering
\caption{\textbf{Trials for the Application User Study}: Each user performed four trials each for two scenarios in a randomized order. The two characters in the scene were either submissive or dominant, as above.}
\begin{tabular}{|c|c|c|c|c|}
\hline
& Trial 1    & Trial 2    & Trial 3    & Trial 4  \\ \hline
Left Character  & HS & HS & HD & HD \\ \hline
Right Character & HS & HD & HS & HD \\ \hline
\end{tabular}
\label{tab:userStudyTrials}
\end{table}

The appearances of the two characters were similar except for the gaits used, and their faces were hidden behind solid cube as in the \blue{validation user study}.

\subsubsection{Questions}
In the \textit{Office} scenario, for each trial, the participants were asked to indicate whether each character was an employee or a boss. They could also choose a third option of \textit{Unsure} if they could not clearly decide. Participants assigned these labels to each character independently. In the \textit{School} scenario, they were asked to indicate whether each character was a student or a teacher, again with the third option of \textit{Unsure}.

\subsection{Discussion}
Table~\ref{tab:applicationResults} presents the percentages of participants that chose student, teacher, and unsure for the \textit{School} scene and the percentages that chose employee, boss, and unsure for the \textit{Office} scene.  

\subsubsection{\blue{School Scenario}}
The results of this scenario suggest that our approach to generating virtual characters with different dominance traits is not enough to create characters that are distinguishable as teacher or student. One possible reason for this result is that the dominance relationship in the vertical dimension between a student and a teacher is not clearly defined and the affective or socio-emotional (horizontal) dimension may be stronger in some cases. The verbal feedback from the participants brought this point to our attention. Another factor that the participants reported as having affected their judgment is that the slow walking gait of the submissive character could be associated with an older teacher.

\subsubsection{\blue{Office Scenario}}
Like in the \textit{School} scene, in the \textit{Office} scene, participants were asked to report whether they can decide if the virtual character is an employee or a boss. 
The results of this scenario suggest that our approach to generating virtual characters with different dominance traits can create characters that are distinguishable as an employee or boss when one character is submissive and the other is dominant, supporting our hypothesis. Although the relationship between submissive traits and the character being an employee was observed, a similar association was not clear when both the characters were dominant. One possible reason for this is that the possibility of two dominant bosses in a single scene was not considered by the participants, leading to the reports of \textit{Unsure}. Overall, the results of our \blue{application user study} indicate that our approach can be used to simulate the vertical dimension of interpersonal social relationships between virtual characters.

\begin{table}
\caption{\textbf{Application User Study Results}: We present the percentages of participants that answered teacher, student, or unsure for the \textit{School} scene (in light gray) and the percentages of participants that answered employee, boss, or unsure for the \textit{Office} scene (in dark gray). For the \textit{Office} scene, our approach to generating virtual characters with different dominance traits is able to create characters that are distinguishable as employees and bosses. Overall, the results of our \blue{application user study} indicate that our approach can be used to simulate the vertical dimension of interpersonal social relationships between virtual characters.}
\centering
\begin{tabular}{|c|c|c|c|c|c|c|c|c|}
\hline
\rowcolor[HTML]{FFFFFF} 
\cellcolor[HTML]{FFFFFF} & \multicolumn{2}{c|}{\cellcolor[HTML]{FFFFFF}Trial 1} & \multicolumn{2}{c|}{\cellcolor[HTML]{FFFFFF}Trial 2} & \multicolumn{2}{c|}{\cellcolor[HTML]{FFFFFF}Trial 3} & \multicolumn{2}{c|}{\cellcolor[HTML]{FFFFFF}Trial 4} \\ \cline{2-9} 
\rowcolor[HTML]{FFFFFF} 
\multirow{-2}{*}{\cellcolor[HTML]{FFFFFF}} & HS & HS & HS & HD & HD & HS & HD & HD \\ \hline
\rowcolor[HTML]{EFEFEF} 
Student & 35 & 40 & 45 & 35 & 30 & 40 & 15 & 20 \\ \hline
\rowcolor[HTML]{EFEFEF} 
Teacher & 25 & 20 & 45 & 55 & 60 & 40 & 45 & 40 \\ \hline
\rowcolor[HTML]{EFEFEF} 
Unsure & 40 & 40 & 10 & 10 & 10 & 20 & 40 & 40 \\ \hline
\rowcolor[HTML]{C0C0C0} 
Employee & 60 & 55 & 65 & 5 & 0 & 85 & 25 & 15 \\ \hline
\rowcolor[HTML]{C0C0C0} 
Boss & 5 & 10 & 10 & 75 & 90 & 0 & 35 & 40 \\ \hline
\rowcolor[HTML]{C0C0C0} 
Unsure & 35 & 35 & 25 & 20 & 10 & 15 & 40 & 45 \\ \hline
\end{tabular}
\label{tab:applicationResults}
\end{table}

\section{Conclusion, Limitations, and Future Work}
We present a novel approach to computing a data-driven mapping between dominance traits and gait features.  This mapping is used at runtime to generate virtual characters with different dominance traits. We validate our approach with a user study in a virtual environment. We present a gait dominance classification method to classify the dominance traits of new input gaits. Our algorithm can classify the dominance of gaits with an accuracy of \blue{\texttildelow73\%}. We also present an application of our approach that simulates the vertical dimension of interpersonal social relationships between virtual characters. In our \blue{application user study}, participants associated roles such as employee or boss to virtual characters based on their dominance traits. To the best of our knowledge, ours is the first approach that models dominance traits for virtual agents at interactive rates. \blue{Additions of realistic human behaviors and interactions have been shown to improve the sense of presence experienced by users when immersed in virtual environments~\cite{blascovich2002immersive,kyriakou2015,guadagno2007virtual}. Our approach to simulating virtual characters with a variety of dominance characteristics can be used to simulate the vertical dimension of the interpersonal relationship between virtual agents. Therefore, our approach is applicable to VR applications like social VR, rehabilitation and PTSD treatments, treatment of crowd phobias, evacuation and urban planning, etc. In addition to these VR applications, our approach can also be used for gaming and simulation applications. Simulating virtual characters with variety of dominance traits can improve the fidelity of character simulations for gaming and training applications.}

Our approach has some limitations. The dominance feature computation is based on the classification of features from visual perception literature. This classification may not be sufficient to capture all observed dominance behaviors. Furthermore, we assume that the motion gait datasets are noise-free, but in practice noise in the joint positions can affect perception. Gaits extracted from RGB videos using state-of-the-art machine learning techniques~\cite{dabral2017structure} contain noise in the joint positions. Our algorithm does not account for these noisy gaits. There are many avenues for future work. In addition to addressing these limitations, we would like to evaluate the performance of our methods for more than two virtual characters. We would also like to identify the importance of trajectories in addition to gait features on the perception of dominance. Furthermore, we would like to combine these dominance traits with other personality traits and evaluate the performance in virtual environments for different applications. 

\ifCLASSOPTIONcompsoc
\else
  \section*{Acknowledgment}
\fi


\ifCLASSOPTIONcaptionsoff
  \newpage
\fi



%

\bibliographystyle{IEEEtran}
\bibliography{IEEEabrv,template}
%

\begin{IEEEbiography}[{\includegraphics[width=1in,height=1.25in,clip,keepaspectratio]{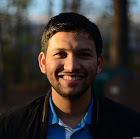}}]{Tanmay Randhavane}
Tanmay Randhavane is a graduate student at the Department of Computer Science, University of North Carolina, Chapel Hill, NC, 27514.
\end{IEEEbiography}

\begin{IEEEbiography}[{\includegraphics[width=1in,height=1.25in,clip,keepaspectratio]{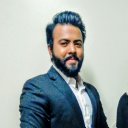}}]{Aniket Bera}
Aniket Bera is a Research Assistant Professor at the Department of Computer Science, University of North Carolina, Chapel Hill, NC, 27514.
\end{IEEEbiography}

\begin{IEEEbiography}[{\includegraphics[width=1in,height=1.25in,clip,keepaspectratio]{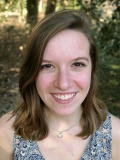}}]{Emily Kubin}
Emily Kubin is a graduate student the Department of Social Psychology, Tilburg University, Tilburg, Netherlands.
\end{IEEEbiography}

\begin{IEEEbiography}[{\includegraphics[width=1in,height=1.25in,clip,keepaspectratio]{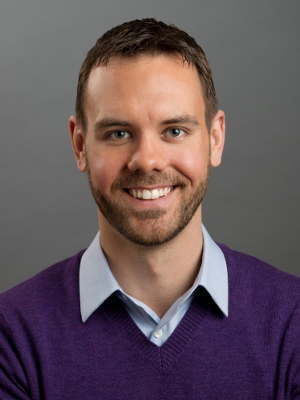}}]{Kurt Gray}
Kurt Gray is a Associate Professor with the Department of Psychology and Neuroscience, University of North Carolina, Chapel Hill, NC, 27514.
\end{IEEEbiography}

\begin{IEEEbiography}[{\includegraphics[width=1in,height=1.25in,clip,keepaspectratio]{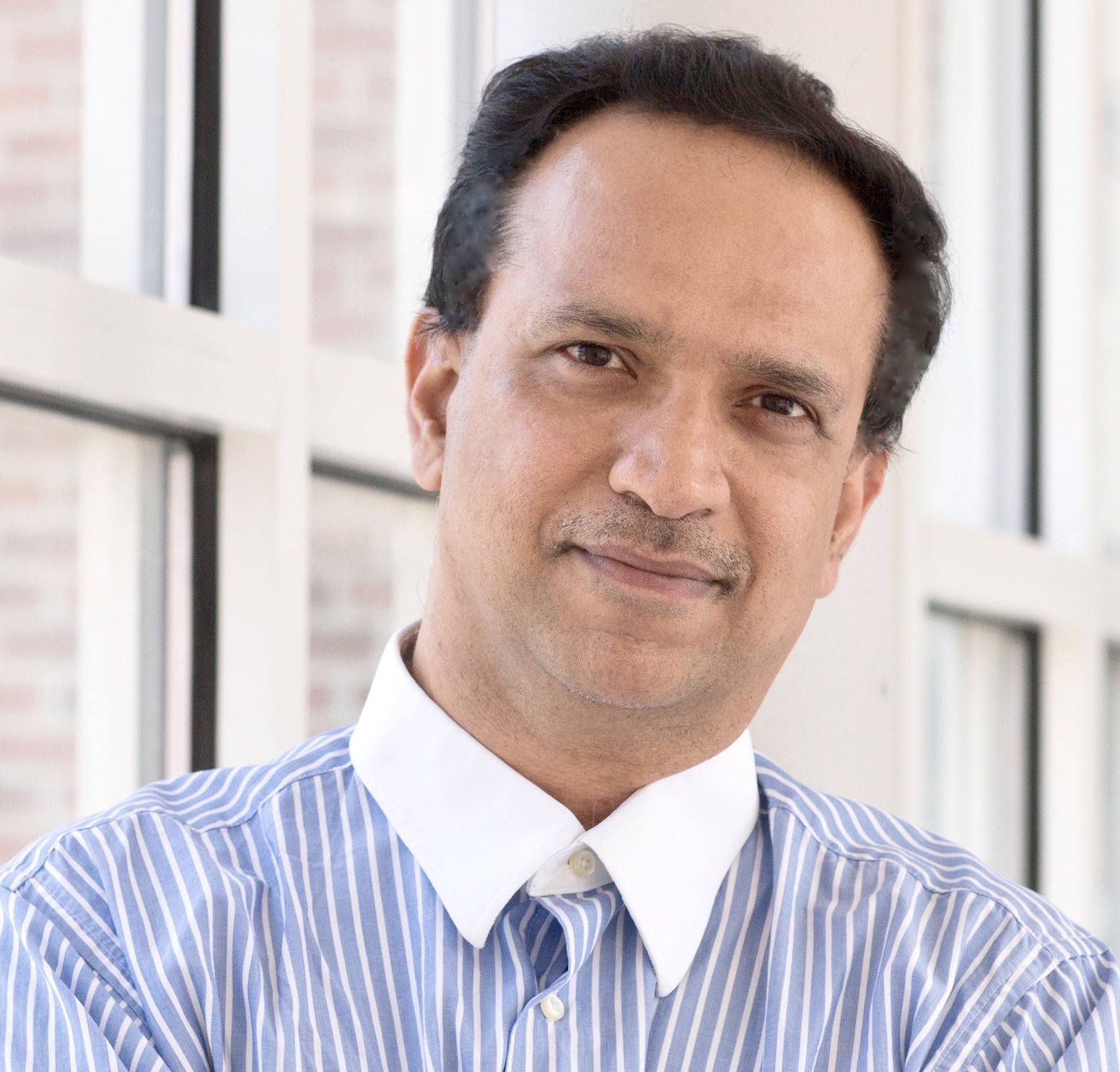}}]{Dinesh Manocha}
Dinesh Manocha is a Paul Chrisman Iribe Chair of Computer Science and Electrical \& Computer Engineering at the University of Maryland at College Park, MD, 20740.
\end{IEEEbiography}






\end{document}